\documentclass{ifacconf}
\usepackage{amsmath}       
\usepackage{float}       
\usepackage{amssymb}     

\usepackage{graphicx}      
\usepackage{natbib}        
\usepackage{xcolor}

\begin{document}
\begin{frontmatter}

\title{Time-delay control for stabilization \\ of the Shapovalov mid-size firm model} 

\thanks[footnoteinfo]{This work has been accepted to the IFAC World Congress 2020.}

\author[First]{T.A.~Alexeeva,} 
\author[Second]{W.A.~Barnett,} 
\author[Third,ipm,fin]{N.V.~Kuznetsov,} 
\author[Third]{T.N.~Mokaev}

\address[First]{St.~Petersburg School of Mathematics, Physics and Computer Science, National Research University Higher School of Economics, Russia (e-mail: tatyanalexeeva@gmail.com)}
\address[Second]{Department of Economics, University of Kansas, USA \\ (e-mail: williamabarnett@gmail.com)}
\address[Third]{Faculty of Mathematics and Mechanics,
\\ St.~Petersburg State University, Russia \\ (e-mails: nkuznetsov239@gmail.com,  tim.mokaev@gmail.com)}
\address[ipm]{Institute for Problems in Mechanical Engineering RAS, Russia}
\address[fin]{Faculty of Information Technology,  University of Jyv\"{a}skyl\"{a}, Finland}

\begin{abstract}                
Control and stabilization of irregular and unstable behavior of dynamic systems (including chaotic processes) are interdisciplinary problems of interest to a variety of scientific fields and applications.
Using the control methods allows improvements in forecasting the dynamics of unstable economic processes and offers opportunities for governments, central banks, and other policy makers to modify the behaviour of the economic system to achieve its best performance.
One effective method for control of chaos and computation of unstable periodic orbits (UPOs) is the unstable delay feedback control (UDFC) approach, suggested by K.~Pyragas. 
This paper proposes the application of the Pyragas' method within framework of economic models. 
We consider this method through the example of the Shapovalov model, by describing the dynamics of a mid-size firm. The results demonstrate that suppressing chaos is capable in the Shapovalov model, using the UDFC method.
\end{abstract}


\end{frontmatter}

\section{Introduction}
Analysis and forecasting of dynamics is one of the main tasks in studies of financial and economic processes. In the framework of this task, a crucial research question is posed to determine the qualitative properties of dynamics (revealing stable and unstable regimes, including chaos). Chaotic behaviour in an economic system is usually an undesirable phenomenon that hinders the accuracy of predictions over long time periods. Many researchers have striven to explain the central features of economic data: irregular and erratic microeconomic and macroeconomic fluctuations, irregular economic growth, structural changes, and overlapping waves of economic development from the point of view of chaos theory, and to demonstrate the complexity and unpredictable behaviour of economic processes via nonlinear dynamic models \citep{Day2-1983,BenhabibD8-1981,Hommes-1995,BrockHomEc-1997,BrockH9-1998,HommesBook12-2006,BarnettSert10-2000,BarnettChProc-1988,BarnettCh:J-1988}.
However, the main goal of economic policy is forecasting the behaviour of economic systems both at macro (country, regions) and micro levels (companies, households).
As John von Neumann correctly noted: ``All stable processes, we shall predict. All unstable processes, we shall control'' \emph{circa 1950} \citep{FangChIsh:Book-2017}.

Controlling at least some economic processes is one of the most challenging tasks facing the economists and politicians responsible for economic policy \citep{Faggini:4-2008,Orlando:4-2006}.
Decision-makers often face the difficult task of dealing with an economic system that behaves in unpredictable ways. In particular, chaotic dynamics can be generated by endogenous nonlinear dynamics without any external influence of the relevant interacting variables \citep{HolystHHW-1996}.
However, if an economic system possesses deterministic characteristics, chaotic behaviour  can be controlled with reliable control methods (see, e.g. \citep{ChenWCh:1-2008}).

Chaos control has been demonstrated in a wide variety of areas, including mechanics, electronics, biology, chemistry and medicine \citep{Rosser:Book-2000,FradkovE-2005,ThomasOSIIS:med-2019,WielandW:10-2005}. 
A number of applications of chaos control methods exist in economic contexts
\citep{MendesMen:2-2005,Neck:15-2009,AmritRA:13ARC-2011,YuCLi:12-2012,NaimzadaP:6-2015,CavalliNP:5-2017,KellettWFGS:14-2019}. 

One of the key problems arising within the frameworks of chaos and control theories is the problem of how to suppress or stabilize chaotic behavior in applied systems. 
These problems were first posed by well-known physicists E.~Ott, S.~Greboggi and J.~York \citep{OttGY-1990} in the 1990s, and remain of great interest. Next, the Pyragas' time-delay feedback control method \citep{Pyragas-1992} have appeared.
The primary mechanism applied to suppress chaos is localization with the help of small perturbations in the system, or introduction of UPO controls embedded in a chaotic attractor.

Holyst et al. \citep{HolystHHW-1996,HolystU-2000,HolystZU:9-2001} showed that applying the Pyragas time-delayed feedback control to the microeconomic Behrens-Feichtinger model can facilitate an easy switch from a chaotic trajectory to a regular periodic orbit and simultaneously improve the system's economic properties. Kopel \citep{Kopel-1997}, using a model of evolutionary market dynamics, demonstrated how chaotic behaviour can be controlled by making small changes in a parameter that is accessible to the decision makers. Bala et al. \citep{BalaMM-1998} proposed to control chaos arising in the context of a ``trial-and-error'' process of exchange economies. Kaas \citep{Kaas-1998} proved that within a macroeconomic disequilibrium model, stationary and simple adaptive policies are not capable of stabilizing efficient steady states and lead to periodic or irregular fluctuations for large sets of policy parameters. Xu et al. \citep{XuBG:18-2002} introduced an approach to detect the unstable periodic orbits (UPOs) pattern from chaotic time series in the Kaldor business cycle model. Salarieh and Alasty \citep{SalariehAl:3-2009} applied the minimum entropy algorithm of chaos control to the Cournot duopoly with different constant marginal costs.


In this work, we show the results of our application of the time-delay feedback control to stabilization and chaos suppression in the Shapovalov mid-size firm model.




\section{Problem Statement}


In the framework of economic models, occurrences of chaotic processes is extremely undesirable. Therefore, the development of effective methods for suppressing chaos and bringing the dynamics of a model to a stable regime by using a small number of corrective operations is an important task. 

Two points turn out to be important for insights into economic policies. Firstly, moving from unstable orbits to other types (for instance, to periodic orbits) on the attractor means choosing different behaviour of the economic system. Secondly, small parameter changes and the presence of many aperiodic orbits can signal resource saving and choosing among different trade-offs of economic policies \citep{BoccalettiGLMM-2000}.
Further, application of control methods to chaotic dynamic systems shows that a government can, in principle, stabilize an unstable equilibrium in a short time by varying income tax rates or government expenditures.
Therefore, using control methods, decision-makers can improve efficiency and induce considerable improvements in the system's economic properties in terms of profits and welfare \citep{Day-1994}.

The application of chaos control methods can be considered both as a way to improve the effectiveness of central bank interventions and as a source of extra monetary policy tools. Inflation targeting frameworks give a central role to transparency and communication of the central bank's views on the economy, its operating procedures, and its expectations.
All of that is easier to communicate with highly regular interest rate dynamics. When deviations from target value are decreased, agents' expectations will be better aligned with what the central bank is trying to anchor them on.  If agents do not fully believe the central bank, more precise forecasts can help build and maintain credibility, making it easier to control expectations. As managing expectations is crucially important for controlling inflation, a central bank's job becomes easier.
Moreover, firms can improve their performance measures using the time-delay feedback control method.

\subsection{Shapovalov Mid-Size Firm Model}

Consider the model of V.I.~Shapovalov proposed in \citep{ShapovalovKBA-2004,ShapovalKaz-2015} which describes the dynamics of a mid-size firm
\begin{equation}
\begin{cases}
\begin{aligned}
&\dot x=-\sigma x+\delta y,\\
&\dot y=\mu x +\mu y-\beta xz,\\
&\dot z=-\gamma z+\alpha xy.\end{aligned} 
\label{sys:Shap6} 
\end{cases}
\end{equation}
Here $\alpha$, \, $\beta$, \, $\sigma$, \, $\delta$, \, $\mu$, \, $\gamma$ are positive parameters, and the variables $x$, $y$, $z$ denote the growth of three main factors of production: the loan amount $x$, fixed capital $y$ and the number of employees $z$ (as an increase in human capital). An increase in the loan amount is proportional to the amount of capital and the size of the loan taken. The coefficient with the variable $y$ is positive according to the premise that, with an increase in capital, the company is more likely to give loans in the lending market; the coefficient for the variable $x$ is negative and indicates the losses that the company incurs when taking a new loan, which is associated with the requirement to pay interest, as well as the fact that the company is less willing to give credit when it has many loan obligations. The capital gain is proportional to the income from the investment of available capital and the loan taken, as well as expenses on labor remuneration and loan repayment. The coefficients for the sum of the variables $x$ and $y$ are positive, since they show a positive effect of investing  in the development of production; the coefficient for the product of the variables $x$ and $z$ is negative, since it indicates the costs of the company. The increase in the number of employees is proportional to the capital, the loan taken and the current number of employees. The coefficient for the product of the variables $x$ and $y$ is positive, based on the assumption that the company may spend part of the amount of capital and the loan taken on attracting additional employees. A negative coefficient for the variable $z$ indicates that the outflow of employees due either to dismissal or on their own initiative should be taken into account.

Coefficients at variables are control parameters: $\alpha$ reflects a combination of factors that contribute to creating a company image that will be attractive to new employees; $\beta$ summarizes factors that influence cost allocation; $\mu$ describes the effectiveness of capital investments (the effects of various taxes should be taken into account); $\gamma$ summarizes factors related to difficulties in obtaining a loan; for example, a high interest rate, etc.


System \eqref{sys:Shap6} can be reduced to a Lorenz-like system
\begin{equation}\label{sys:shapovalov_lor}
  \begin{cases}
\dot x= - c (x + y), \\
\dot  y=r x + y - x z, \,\,\mbox{where} \,\, c = \frac{\sigma}{\mu} , r=\frac{\delta}{\sigma}, b = \frac{\gamma}{\mu},\\
\dot z=-b z + xy, 
  \end{cases}
\end{equation}

using the following coordinate transformation
\begin{equation}
(x, y, z) \rightarrow  \left(\frac{\mu}{\sqrt{\alpha \beta}} x,\, \frac{\mu \sigma}{\delta \sqrt{\alpha \beta}} y, \, \frac{\mu \sigma}{\delta \beta} z\right), \, t \rightarrow \frac{t}{\mu}. 
\label{eq:TrfShLor} 
\end{equation}
System \eqref{sys:shapovalov_lor} differs from the classical Lorenz system \citep{Lorenz-1963} in the sign of the coefficient at $y$ in the second equation, which is 1 here, while in the Lorenz system this coefficient is -1.

Accordingly, the inverse transformation
\begin{equation}
(x, y, z) \rightarrow  \left(\frac{\sqrt{\alpha \beta}}{\mu} x, \frac{r \sqrt{\alpha \beta}}{\mu} y, \frac{r \beta}{\mu} z\right),\, t \rightarrow \mu t
\label{eq:InvtrfLorSh} 
\end{equation}
reduces system \eqref{sys:shapovalov_lor} to system \eqref{sys:Shap6} with coefficients $\sigma = c \mu, \delta = r c \mu, \gamma = b \mu$.

As part of the study of system \eqref{sys:Shap6}, \citep{ShapovalovKBA-2004, GurinaD-2010,ShapovalKaz-2015} formulated the task of nonlinear analysis of the system and its limit dynamics in order to forecast and control behaviour in the Shapovalov model \eqref{sys:Shap6}. Revealing unstable periodic orbits and chaos suppression in the Shapovalov model are challenging tasks.

\subsection{Stabilization of Periodic Orbits in the Shapovalov Model via Time-Delay Feedback Control}

One effective method among others for the stabilization of UPOs
is the \emph{delay feedback control} (DFC) approach,
suggested by Pyragas \citep{Pyragas-1992}
(see discussions of its advantages, limitations and modifications
in \citep{KuznetsovLS-2015-IFAC,ChenY-1999,LehnertHFGFS-2011,HootonA-2012}).
This approach allows Pyragas and his progeny
to stabilize and study UPOs in various chaotic dynamic systems.
Below, a modification of the classical DFC
technique, termed the unstable delayed feedback control (UDFC)~\citep{Pyragas-2001}, is used.

We rewrite system \eqref{sys:shapovalov_lor} in a general form
\begin{equation}\label{eq:sys_gen}
  \dot{u} = f(u).
\end{equation}
Let $u^{\rm upo}(t,u^{\rm upo_1}_0)$ be its UPO with period $\tau > 0$,
$u^{\rm upo}(t - \tau,u^{\rm upo_1}_0) = u^{\rm upo}(t,u^{\rm upo_1}_0)$,
and initial condition $u^{\rm upo_1}_0=u^{\rm upo}(0,u^{\rm upo_1}_0)$.
To compute the UPO, we add the UDFC in the following form:
\begin{equation}\label{eq:closed_loop_syst}
 \begin{aligned}
\dot{u}(t) &= f(u(t)) + K B \, \big[F_N(t) + w(t)\big], \\
\dot{w}(t) &= \lambda_c^0 w(t) + (\lambda_c^0 - \lambda_c^\infty) F_N(t), \\
F_N(t) &= C^*u(t) - (1\!-\!R) \sum_{k=1}^N R^{k-1} C^*u(t - k T),
  \end{aligned}
\end{equation}
where
$0 \leq R < 1$ is an extended DFC parameter,
$N = 1,2,\ldots,\infty$ defines the number of previous states involved in
delayed feedback function $F_N(t)$, 
$\lambda_c^0 > 0$, and $\lambda_c^\infty < 0$ are
additional unstable degree of freedom parameters,
$B, C$ are vectors and $K > 0$ is a feedback gain.
For the initial condition $u^{\rm upo_1}_0$ and $T = \tau$ we have
\[
  F_N(t) \equiv 0, \ w(t) \equiv 0,
\]
and, thus, the solution of system \eqref{eq:closed_loop_syst}
coincides with the periodic solution
of the initial system \eqref{eq:sys_gen}.

\begin{figure}[ht]
\center{\includegraphics[width=8.4cm]{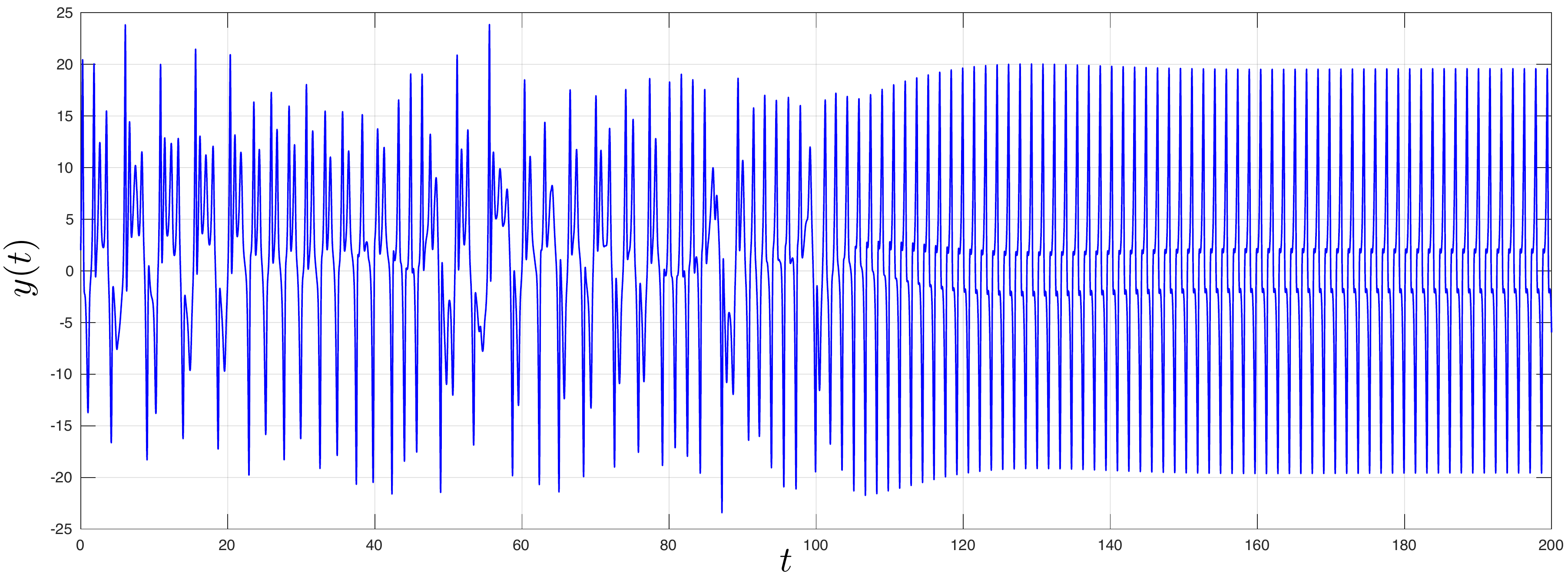}} \\ a) 
\center{\includegraphics[width=8.4cm]{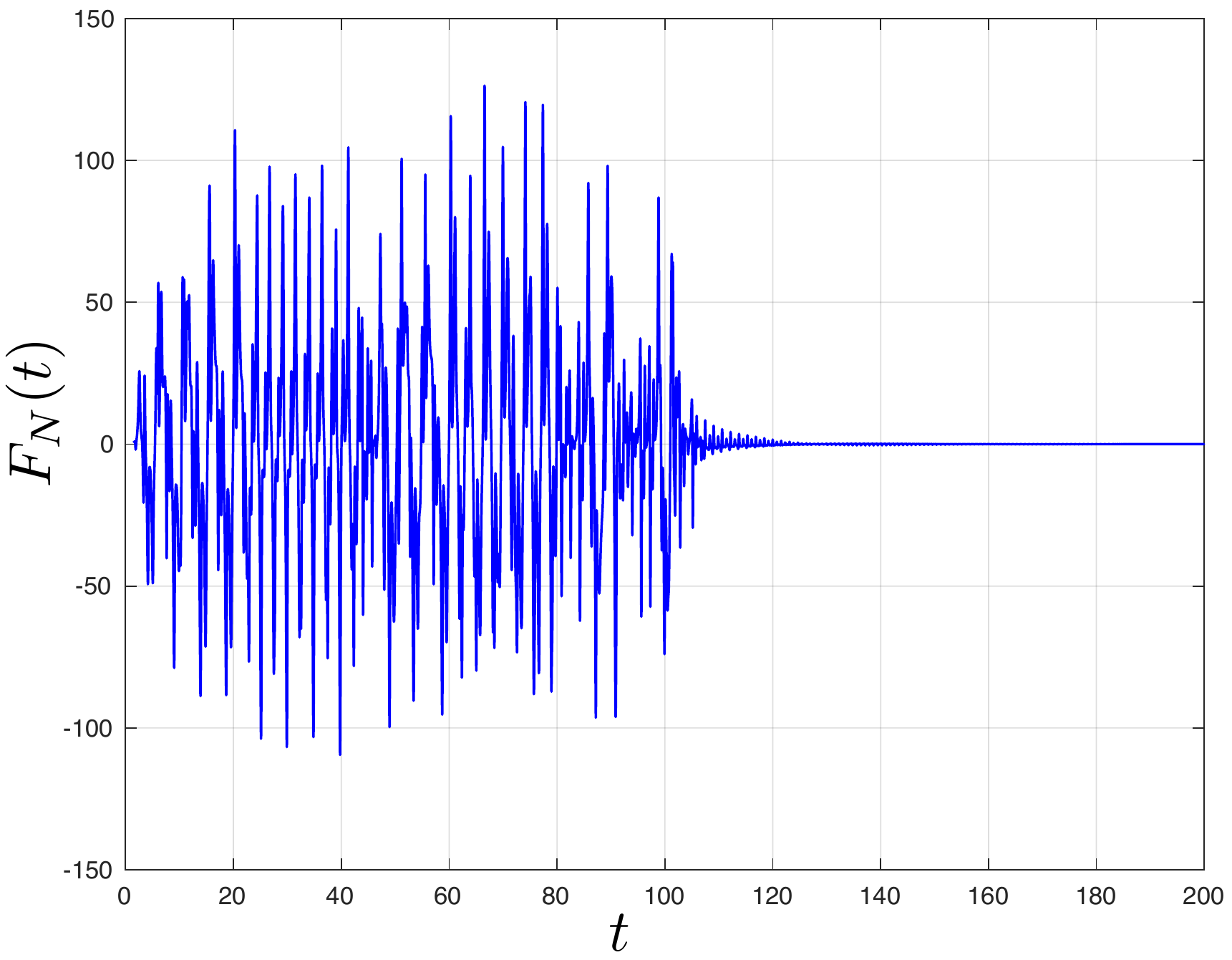}} \\ b) 
 \center{\includegraphics[width=8.4cm]{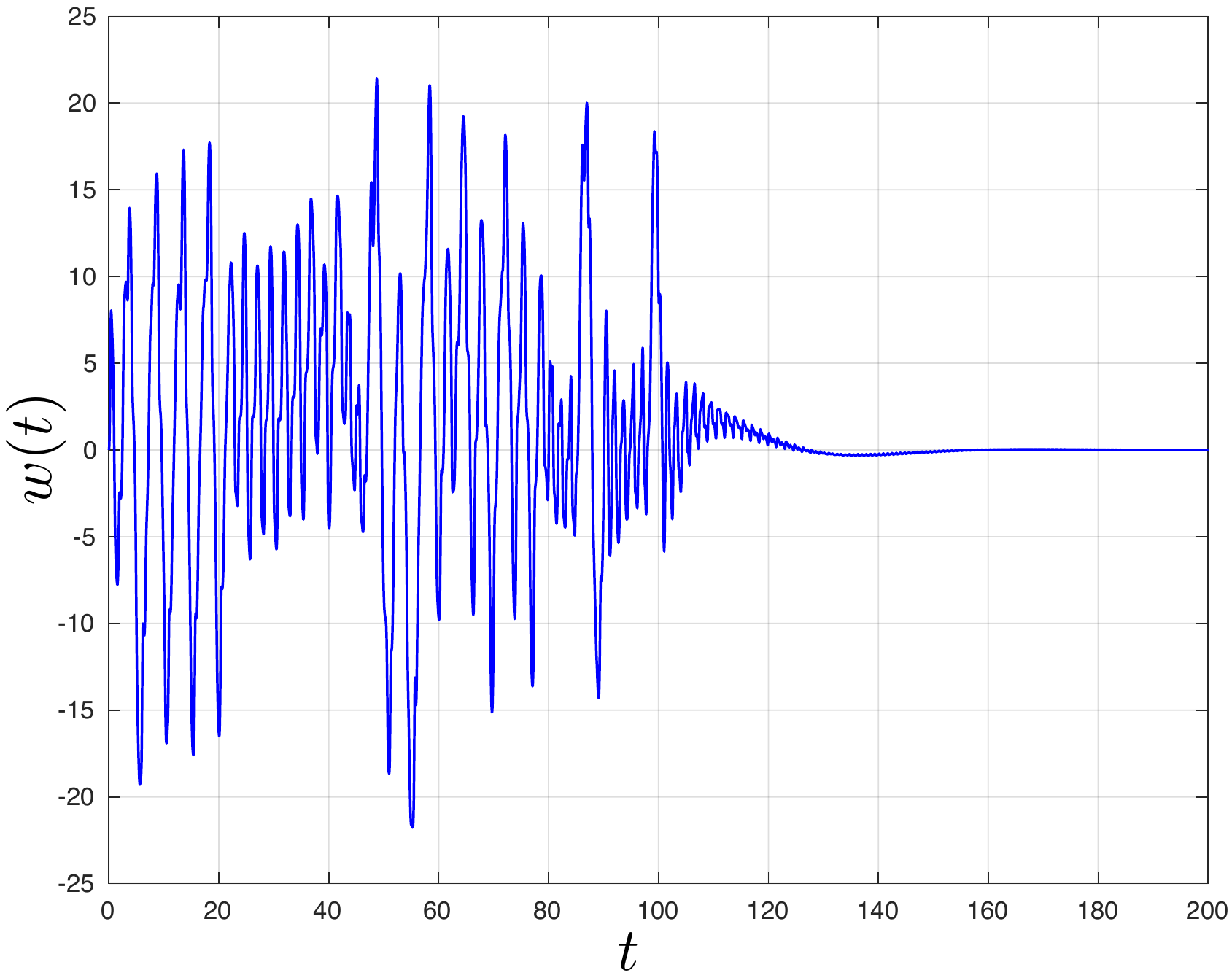}} \\ c) 
  \caption{Chaos suppression in system \eqref{sys:shapovalov_lor}
   with parameters $r = 28$, $c = 10$, $b = 8/3$ using the UDFC method:
   a) time evolution of the measurable variable $y(t) = C^*u(t)$; b) time-delayed feedback control $F_N(t)$; c) an additional unstable degree of freedom $w(t)$.
   }
   \label{fig:shapovalov:stab:w}
 \end{figure}

\begin{figure}[ht]
\begin{center}
   \includegraphics[width=8.4cm]{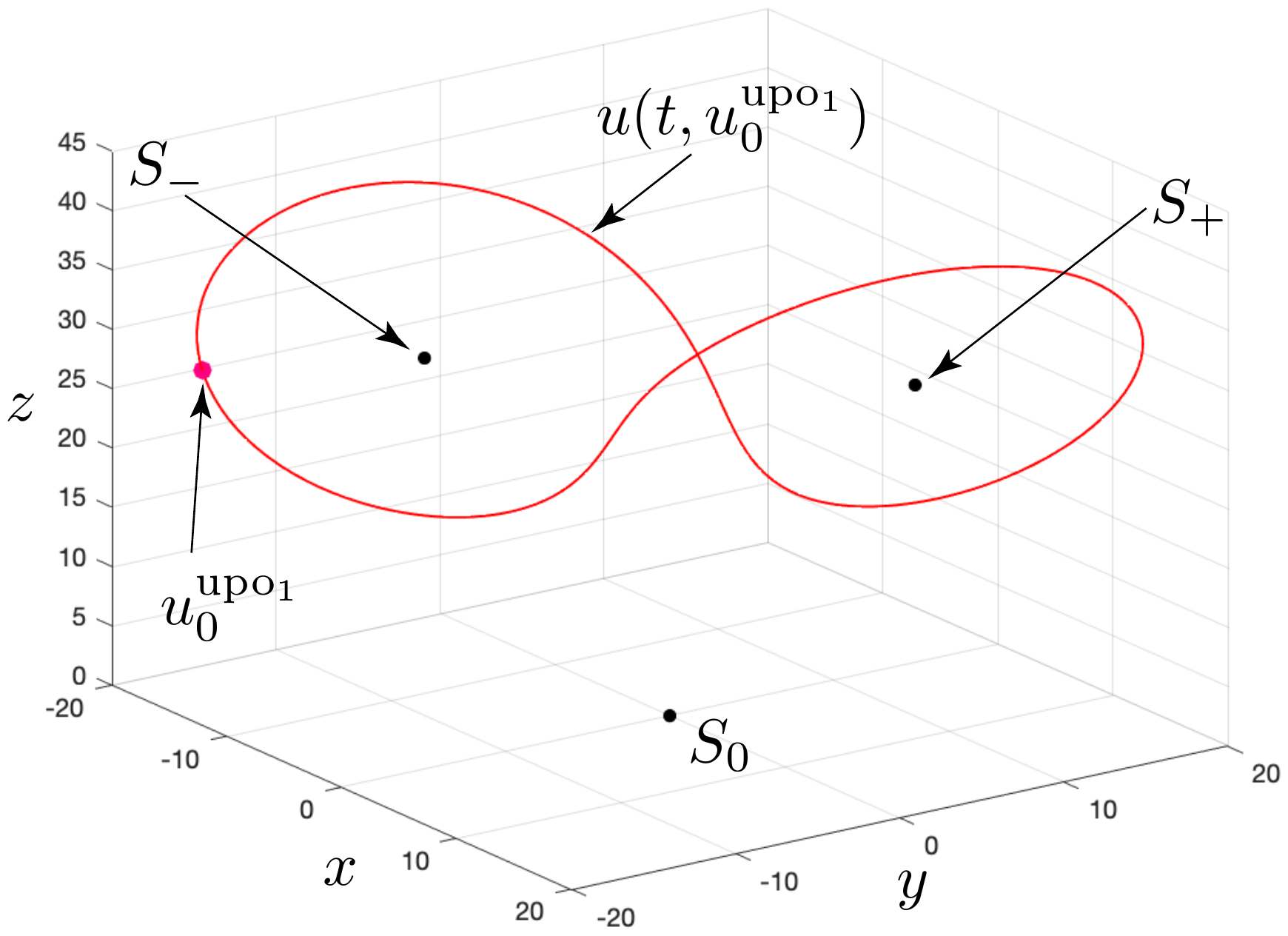}
   \caption{Period-1 UPO $u^{\rm upo_1}(t)$ (red, period $\tau_1 = 1.28701$)
    stabilized using the UDFC method in system \eqref{sys:shapovalov_lor} with parameters set at $r = 28$, $c = 10$, $b = 8/3$.
    }
   \label{fig:shapovalov:upo}
\end{center}
\end{figure}

For system~\eqref{sys:shapovalov_lor}, for example, 
with "Lorenz-like" parameters $r = 28$, $c = 10$, $b = 8/3$
using~\eqref{eq:closed_loop_syst} with
$B^{*} = \left(0, 1, 0\right)$, $C^* = \left(0, 1, 0\right)$,
$R = 0.4$, $N =100$, $K = 0.5$, $\lambda^0_c = 0.01$, $\lambda^\infty_c = -0.5$,
one can stabilize a period-1 UPO $u^{\rm upo_1}(t,u_0)$
with period $\tau_1 = 1.28701$
from the initial point $u_0 = (1, -1, 1)$, $w_0 = 0$ on the time interval $[0, 200]$
(see Fig.~\ref{fig:shapovalov:stab:w} and Fig.~\ref{fig:shapovalov:upo}).

The existence of the UPO obtained can be verified by various other
numerical (see, e.g. \citep{Viswanath-2001,Budanov-2018,PchelintsevPY-2019-LorenzUpo})
and computer-assisted (see, e.g. \citep{GaliasT-2008-LorenzUpo,BarrioDT-2015-LorenzUpo})
approaches.
However, the Pyragas procedure, in general, is more convenient for numerical
stabilization and visualization of UPOs.

\begin{figure}[ht]
\begin{center}
     \includegraphics[width=8.4cm]{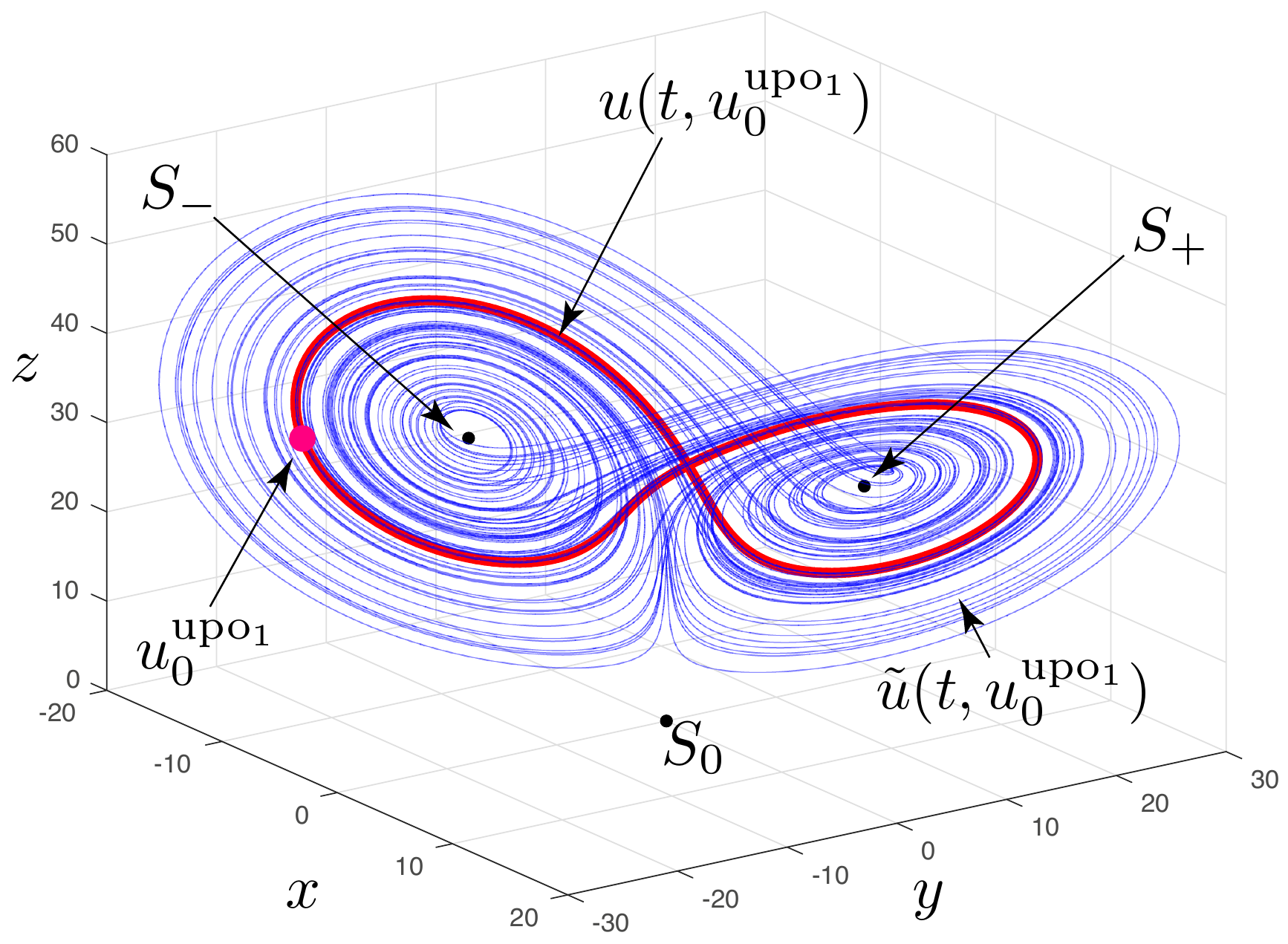}
    \caption{Period-1 UPO $u^{\rm upo_1}(t)$ (red, period $\tau_1 = 1.28701$)
    stabilized using the UDFC method,
    and pseudo-trajectory $\tilde{u}(t, u^{\rm upo_1}_0)$ (blue, $t \in [0,100]$) in system \eqref{sys:shapovalov_lor} with parameters set at $r = 28$, $c = 10$, $b = 8/3
    $. 
    }
    \label{fig:shapovalov:upo1attr}
 \end{center}
 \end{figure}

For the initial point $u^{\rm upo_1}_0 \approx (-13.3549, -19.1419, 29.1187)$
on the UPO $u^{\rm upo_{1}}(t) = u(t, u^{\rm upo_1}_0)$
we numerically compute the trajectory
of system \eqref{eq:closed_loop_syst} without the stabilization (i.e. with $K = 0$)
on the time interval $[0,T=100]$ (see Fig.~\ref{fig:shapovalov:upo1attr}).
We denote it by $\tilde{u}(t, u^{\rm upo_1}_0)$
to distinguish this pseudo-trajectory from the periodic orbit $u(t, u^{\rm upo_1}_0)$.
On the initial small time interval $[0,T_1 \approx 10]$,
even without the control,
the obtained trajectory $\tilde{u}(t, u^{\rm upo_1}_0)$
approximately traces the ''true'' trajectory (periodic orbit) $u(t, u^{\rm upo_1}_0)$.
But for $t > T_1$, without a control,
the pseudo-trajectory $\tilde{u}(t, u^{\rm upo_1}_0)$
diverges from $u(t, u^{\rm upo_1}_0)$ and
visualize a local chaotic attractor~$\mathcal{A}$.

Concerning time required for integration, while the time series obtained from a \emph{physical experiment} are assumed to be reliable on the overall time interval considered,
time series produced using the integration 
\emph{mathematical dynamic model} can be reliable on a limited time interval only
due to computational errors
(caused by finite precision arithmetic and numerical integration of ODE).
Thus, in general, the closeness of the real trajectory $u(t,u_0)$
and the corresponding pseudo-trajectory $\tilde u(t,u_0)$
calculated numerically can be guaranteed on a limited short time interval only.

\begin{figure}[ht]
\begin{center}
     \includegraphics[width=8.4cm]{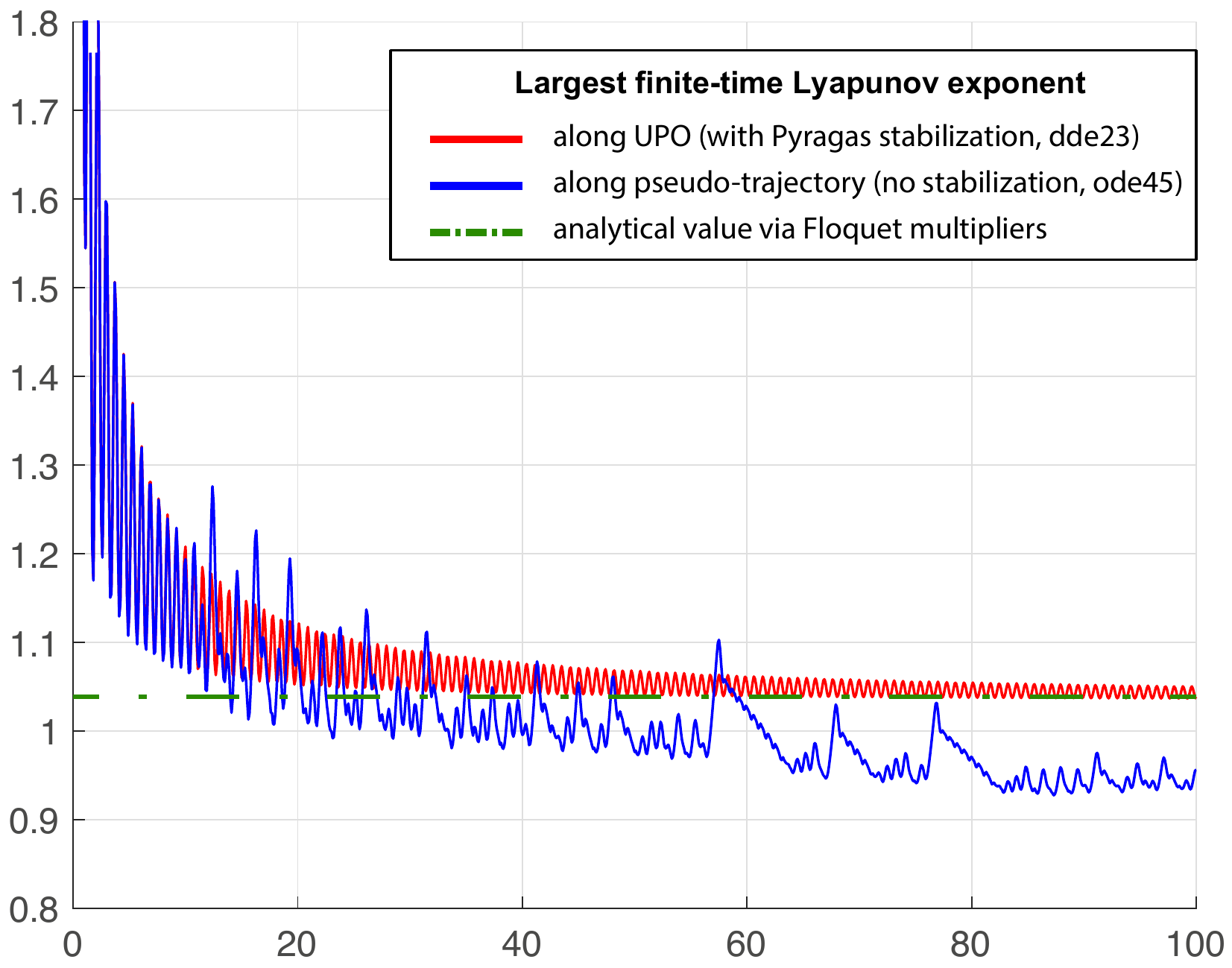}
    \caption{Period-1 UPO $u^{\rm upo_1}(t)$ (red, period $\tau_1 = 1.28701$)
    stabilized using the UDFC method,
    pseudo-trajectory $\tilde{u}(t, u^{\rm upo_1}_0)$ (blue), and the analytical value ${\rm LE}_1(u_{0}^{upo_1})$ (green) for $t \in [0,100]$ in system \eqref{sys:shapovalov_lor} with parameters set at $r = 28$, $c = 10$, $b = 8/3$.
    }
    \label{fig:shapovalov:LE}
 \end{center}
 \end{figure}

The obtained values of the largest \emph{finite-time Lyapunov exponent} (FTLE) ${\rm LE}_1(t,u_{0}^{upo_1})$ computed along the stabilized UPO $u(t, u^{\rm upo_1}_0)$ and the trajectory without stabilization $\tilde{u}(t, u^{\rm upo_1}_0)$ gives us the following  results. On the initial part of the time interval $[0,T_1 \approx 10]$, one can indicate the coincidence of these values with a sufficiently high accuracy. 
After $t > T_2\approx 40$ the difference in values becomes significant and the corresponding graphs diverge in such a way that the graph corresponding to the unstabilized trajectory is lower than the parts of the graphs corresponding to the UPO and the analytical value largest Lyapunov exponent: ${\rm LE}_1(u_{0}^{upo_1})=1.038532560368980$, computed via Floquet multipliers (see Fig.~\ref{fig:shapovalov:LE}). 
The numerical integration of trajectories via approximate methods is strongly influenced by the round-off and truncation errors which in general accumulate over a large time interval and do not allow tracking the "true" trajectory without the significant increase of the precision of the floating-point representation. 
These results are in close agreement with 
rigorous analyses of the time interval choices for reliable numerical computation of trajectories
for the Lorenz system:
the time interval for reliable computation 
with 16~significant digits and error $10^{-4}$
is estimated as $[0, 36]$, with error $10^{-8}$
 estimated as $[0, 26]$ (see \citep{KehletL-2013,KehletL-2017}),
and reliable computation for a longer time interval, e.g. $[0,10000]$ in Liao and Wang \citep{LiaoW-2014},
is a challenging task that requires significant increase of the precision of the floating-point
representation and the use of supercomputers.
Analytical aspects of this problem are related to
the shadowing theory. 


 \begin{figure}[ht]
\begin{center}
     \includegraphics[width=8.4cm]{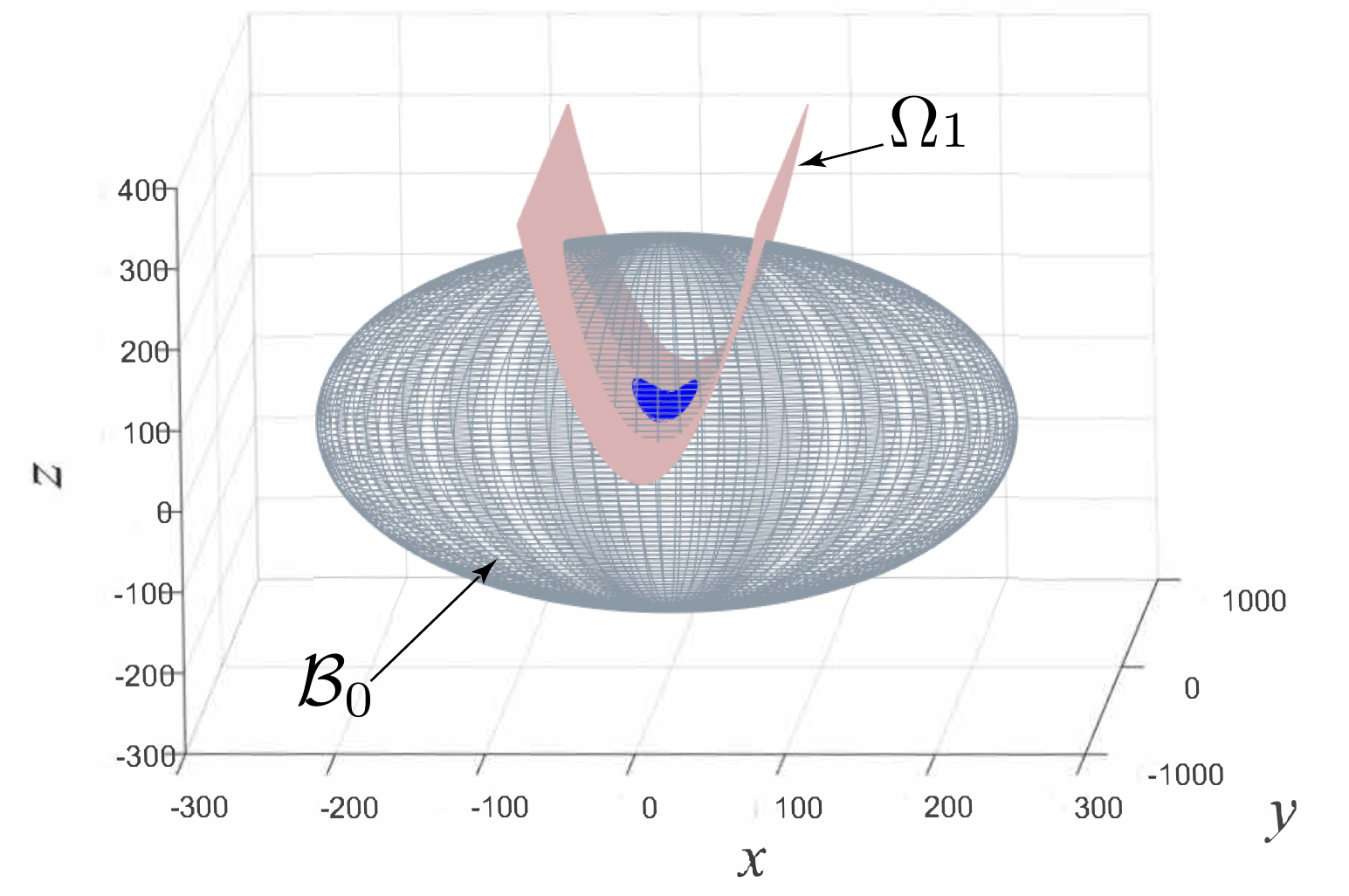}
    \caption{Localization of the chaotic attractor (blue) of system~\eqref{sys:shapovalov_lor}
  with parameters $r = 28$, $c = 10$, $b = 8/3$ by the
  absorbing set $\mathcal{B}=\mathcal{B}_0 \bigcap \Omega_1$, where $\mathcal{B}_0$ is the ellipsoid (gray), $\Omega_1$ is the parabolic cylinder (brown).}
    \label{fig:shapovalov:absell}
 \end{center}
 \end{figure}



\section{Conclusion}

Control of chaos remains an area of intensive research. Reliable forecasting of the dynamics of nonlinear systems with chaotic behaviour is a challenging task. It can be solved in several ways.  For example, by localizing a chaotic attractor, while obtaining a rough forecasting, or by introducing control of UPOs embedded in a chaotic attractor, thereby making the behavior of the system predictable for given values of its parameters. Using the example of a three-dimensional dynamic system, we examine the application of the Pyragas time-delay feedback control technique for suppressing chaos in a mid-size firm model. From the essential economic point of view of the task this result means that decision-makers can not only control the behaviour of an economic system, but also choose preferred periodic solutions; that is, increase the predictability of the behavior of the economy. A similar procedure can be performed for other UPOs and model parameters. Using analytical methods \citep{LeonovKM-2015-EPJST,Kuznetsov-2016-PLA,KuznetsovAL-2016,LeonovKKK-2016-CNSCS}, two important cases of predictable dynamics can be distinguished in the parameter space of the Shapovalov model. First, by constructing the Lyapunov function, it is possible to indicate the region of parameters for which all the trajectories of the system eventually fall into a limited region of the phase space (see Fig.~\ref{fig:shapovalov:absell}). Second, by analyzing the matrix of the first approximation, we can ascertain a range of parameters for which all trajectories tend to a stationary set (which consists of all stable and unstable equilibrium states).

\begin{ack}
The work is supported by the Leading Scientific Schools of Russia project NSh-2624.2020.1 and the Russian Science Foundation project 19-41-02002 (section 2.2).  
This work is done under the umbrella of the Institute for Nonlinear Dynamical Inference at the International Center for Emerging Markets Research (http://icemr.ru/institute-for-nonlinear-dynamical-inference/).
\end{ack}

                                       

\begin{thebibliography}{57}
\providecommand{\natexlab}[1]{#1}
\providecommand{\url}[1]{\texttt{#1}}
\providecommand{\urlprefix}{URL }
\expandafter\ifx\csname urlstyle\endcsname\relax
  \providecommand{\doi}[1]{doi:\discretionary{}{}{}#1}\else
  \providecommand{\doi}{doi:\discretionary{}{}{}\begingroup
  \urlstyle{rm}\Url}\fi

\bibitem[{Amrit and Angeli(2011)}]{AmritRA:13ARC-2011}
Amrit, R., R.J.B. and Angeli, D. (2011).
\newblock Economic optimization using model predictive control with a terminal
  cost.
\newblock \emph{Annual Reviews in Control}, 35, 178--186.

\bibitem[{Bala et~al.(1998)Bala, Majumdar, and Mitra}]{BalaMM-1998}
Bala, V., Majumdar, M., and Mitra, T. (1998).
\newblock A note on controlling a chaotic tatonnement.
\newblock \emph{Journal of Economic Behavior and Organization}, 3, 411--420.

\bibitem[{Barnett and Serletis(2000)}]{BarnettSert10-2000}
Barnett, W. and Serletis, A. (2000).
\newblock Martingales, nonlinearity, and chaos.
\newblock \emph{Journal of Economic Dynamics and Control}, 24, 703--724.

\bibitem[{Barnett and Chen(1988{\natexlab{a}})}]{BarnettChProc-1988}
Barnett, W. and Chen, P. (1988{\natexlab{a}}).
\newblock The aggregation theoretic monetary aggregates are chaotic and have
  strange attractors: An econometric application of mathematical of chaos.
\emph{Proc. 3rd International Symposium on Economic Theory and Econometrics}.

\bibitem[{Barnett and Chen(1988{\natexlab{b}})}]{BarnettCh:J-1988}
Barnett, W. and Chen, P. (1988{\natexlab{b}}).
\newblock Deterministic chaotic and fractal attractors as tools for
  nonparametric dynamical econometric inference: With an applications to
  divisia monetary aggregates.
\newblock \emph{Mathematical Computational Modelling}, 10, 275--296.

\bibitem[{Barrio et~al.(2015)Barrio, Dena, and
  Tucker}]{BarrioDT-2015-LorenzUpo}
Barrio, R., Dena, A., and Tucker, W. (2015).
\newblock A database of rigorous and high-precision periodic orbits of the
  {L}orenz model.
\newblock \emph{Computer Physics Communications}, 194, 76--83.

\bibitem[{Benhabib and Day(1981)}]{BenhabibD8-1981}
Benhabib, J. and Day, R. (1981).
\newblock Rational choice and erratic behaviour.
\newblock \emph{Review of Economic Studies}, XLVIII, 459--471.

\bibitem[{Boccaletti et~al.(2000)Boccaletti, Grebogi, Lai, Mancini, and
  Maza}]{BoccalettiGLMM-2000}
Boccaletti, S., Grebogi, C., Lai, Y.C., Mancini, H., and Maza, D. (2000).
\newblock The control of chaos: Theory and applications.
\newblock \emph{Physics Reports}, 329(3), 103--197.

\bibitem[{Brock and Hommes(1997)}]{BrockHomEc-1997}
Brock, W.A. and Hommes, C.H. (1997).
\newblock A rational route to randomness.
\newblock \emph{Econometrica 65, 1059–1095 (1997).}, 65, 1059--1095.

\bibitem[{Brock and Hommes(1998)}]{BrockH9-1998}
Brock, W. and Hommes, C. (1998).
\newblock Heterogeneous beliefs and routes to chaos in a simple asset pricing
  model.
\newblock \emph{Journal of Economic Dynamics and Control}, 22, 1235--1274.

\bibitem[{Budanov(2018)}]{Budanov-2018}
Budanov, V. (2018).
\newblock Undefined frequencies method.
\newblock \emph{Fundam. Prikl. Mat.}, 22, 59--71.
\newblock (in Russian).

\bibitem[{Cavalli and Pecora(2017)}]{CavalliNP:5-2017}
Cavalli, F., N.A. and Pecora, N. (2017).
\newblock Real and financial market interactions in a multiplier-accelerator
  model: Nonlinear dynamics, multistability and stylized facts.
\newblock \emph{Chaos}, 27, 103--120.

\bibitem[{Chen and Yu(1999)}]{ChenY-1999}
Chen, G. and Yu, X. (1999).
\newblock On time-delayed feedback control of chaotic systems.
\newblock \emph{IEEE Transactions on Circuits and Systems I
}, 46(6), 767--772.

\bibitem[{Chen(2008)}]{ChenWCh:1-2008}
Chen, W.C. (2008).
\newblock Dynamics and control of a financial system with time-delayed
  feedbacks.
\newblock \emph{Chaos, Solitons \& Fractals}, 37, 1198--1207.

\bibitem[{Day(1983)}]{Day2-1983}
Day, R.H. (1983).
\newblock The emergence of chaos from classical economic growth.
\newblock \emph{The Quarterly Journal of Economics}, 98(2), 201--213.

\bibitem[{Day(1994)}]{Day-1994}
Day, R.H. (1994).
\newblock \emph{Business Cycles: Theory and Empirical Methods}, vol.~41,
  ch. Business cycles, Fiscal policy and budget deficits, 113--143.
\newblock Springer, Dordrecht.

\bibitem[{Faggini(2008)}]{Faggini:4-2008}
Faggini, M. (2008).
\newblock Analysis of economic fluctuations: A contributions from chaos
  theory.
\newblock In M.~Sibillo and C.~Perna (eds.), \emph{Mathematical and Statistical
  Methods for Insurance and Finance, Springer, Berlin}, 107--112.

\bibitem[{Fang et~al.(2017)Fang, Chen, and Ishii}]{FangChIsh:Book-2017}
Fang, S., Chen, J., and Ishii, H. (2017).
\newblock \emph{Towards Integrating Control and Information Theories. From
  Information-Theoretic Measures to Control Performance Limitations}.
\newblock Springer International Publishing.

\bibitem[{Fradkov and Evans(2005)}]{FradkovE-2005}
Fradkov, A. and Evans, R. (2005).
\newblock Control of chaos: {M}ethods and applications in engineering.
\newblock \emph{Annual Reviews in Control}, 29(1), 33--56.

\bibitem[{Galias and Tucker(2008)}]{GaliasT-2008-LorenzUpo}
Galias, Z. and Tucker, W. (2008).
\newblock Short periodic orbits for the {L}orenz system.
\newblock In \emph{2008 International Conference on Signals and Electronic
  Systems}, 285--288. IEEE.

\bibitem[{Gurina and Dorofeev(2010)}]{GurinaD-2010}
Gurina, T. and Dorofeev, I. (2010).
\newblock Suschestvovanyie gomoklinicheskoy babochki v modely sredney firmy (in
  {R}ussian).
\newblock \emph{Dinamicheskyie sistemy}, 28, 63--68.

\bibitem[{Holyst et~al.(1996)Holyst, Hagel, Haag, and
  Weidlich}]{HolystHHW-1996}
Holyst, J., Hagel, T., Haag, G., and Weidlich, W. (1996).
\newblock How to control a chaotic economy?
\newblock \emph{Journal of Evolutionary Economics}, 6, 31--42.

\bibitem[{Holyst and Urbanowicz(2000)}]{HolystU-2000}
Holyst, J. and Urbanowicz, K. (2000).
\newblock Chaos control in economical model by time-delayed feedback control.
\newblock \emph{Physica A}, 287, 587--598.

\bibitem[{Holyst et~al.(2001)Holyst, Zebrowska, and
  Urbanowicz}]{HolystZU:9-2001}
Holyst, J., Zebrowska, M., and Urbanowicz, K. (2001).
\newblock Observations of deterministic chaos in financial time series by
  recurrence plots, can one control chaotic economy?
\newblock \emph{Physics of Condensed Matter}, 20(4), 531--535.

\bibitem[{Hommes(1995)}]{Hommes-1995}
Hommes, C. (1995).
\newblock A reconsideration of {H}icks’ non-linear trade cycle model.
\newblock \emph{Structural change and economic dynamics}, 6, 435--459.

\bibitem[{Hommes(2006)}]{HommesBook12-2006}
Hommes, C. (2006).
\newblock \emph{Handbook of Computational Economics}, volume~2, chapter~23,
  1109--1186.
\newblock Elsevier.

\bibitem[{Hooton and Amann(2012)}]{HootonA-2012}
Hooton, E. and Amann, A. (2012).
\newblock Analytical limitation for time-delayed feedback control in autonomous
  systems.
\newblock \emph{Phys. Rev. Lett.}, 109, 154101.

\bibitem[{Kaas(1998)}]{Kaas-1998}
Kaas, L. (1998).
\newblock Stabilizing chaos in a dynamic macroeconomic model.
\newblock \emph{Journal of Economic Behavior and Organization}, 33, 333--362.

\bibitem[{Kehlet and Logg(2017)}]{KehletL-2017}
Kehlet, B. and Logg, A. (2017).
\newblock \emph{A posteriori} error analysis of round-off errors in the numerical
  solution of ordinary differential equations.
\newblock \emph{Numerical Algorithms}, 76(1), 191--210.

\bibitem[{Kehlet and Logg(2013)}]{KehletL-2013}
Kehlet, B. and Logg, A. (2013).
\newblock Quantifying the computability of the {L}orenz system using \emph{a
  posteriori} analysis.
\newblock In \emph{Proceedings of the VI Int. conf. on Adaptive Modeling and
  Simulation (ADMOS 2013)}.

\bibitem[{Kellett et~al.(2019)Kellett, Weller, Faulwasser, Grüne, and
  Semmler}]{KellettWFGS:14-2019}
Kellett, C., Weller, S., Faulwasser, T., Grüne, L., and Semmler, W. (2019).
\newblock Feedback, dynamics, and optimal control in climate economics.
\newblock \emph{Annual Reviews in Control}, 47, 7--20.

\bibitem[{Kopel(1997)}]{Kopel-1997}
Kopel, M. (1997).
\newblock Improving the performance of an economic system: controlling chaos.
\newblock \emph{Journal of Evolutionary Economics}, 7, 269--289.

\bibitem[{Kuznetsov(2016)}]{Kuznetsov-2016-PLA}
Kuznetsov, N. (2016).
\newblock The {L}yapunov dimension and its estimation via the {L}eonov method.
\newblock \emph{Physics Letters A}, 380(25-26), 2142--2149.

\bibitem[{Kuznetsov et~al.(2016)Kuznetsov, Alexeeva, and
  Leonov}]{KuznetsovAL-2016}
Kuznetsov, N., Alexeeva, T., and Leonov, G. (2016).
\newblock Invariance of {L}yapunov exponents and {L}yapunov dimension for
  regular and irregular linearizations.
\newblock \emph{Nonlinear Dynamics}, 85(1), 195--201.

\bibitem[{Kuznetsov et~al.(2015)Kuznetsov, Leonov, and
  Shumafov}]{KuznetsovLS-2015-IFAC}
Kuznetsov, N., Leonov, G., and Shumafov, M. (2015).
\newblock A short survey on {P}yragas time-delay feedback stabilization and odd
  number limitation.
\newblock \emph{IFAC-PapersOnLine}, 48(11), 706--709.
\newblock \doi{10.1016/j.ifacol.2015.09.271}.

\bibitem[{Lehnert et~al.(2011)Lehnert, H{\"o}vel, Flunkert, Guzenko, Fradkov,
  and Sch{\"o}ll}]{LehnertHFGFS-2011}
Lehnert, J., H{\"o}vel, P., Flunkert, V., Guzenko, P., Fradkov, A., and
  Sch{\"o}ll, E. (2011).
\newblock Adaptive tuning of feedback gain in time-delayed feedback control.
\newblock \emph{Chaos}
  21(4), 043111.

\bibitem[{Leonov et~al.(2016)Leonov, Kuznetsov, Korzhemanova, and
  Kusakin}]{LeonovKKK-2016-CNSCS}
Leonov, G., Kuznetsov, N., Korzhemanova, N., and Kusakin, D. (2016).
\newblock {L}yapunov dimension formula for the global attractor of the {L}orenz
  system.
\newblock \emph{Commun Nonlinear Sci Numer Simulat},
  41, 84--103.

\bibitem[{Leonov et~al.(2015)Leonov, Kuznetsov, and
  Mokaev}]{LeonovKM-2015-EPJST}
Leonov, G., Kuznetsov, N., and Mokaev, T. (2015).
\newblock Homoclinic orbits, and self-excited and hidden attractors in a
  {L}orenz-like system describing convective fluid motion.
\newblock \emph{The European Physical Journal Special Topics}, 224(8),
  1421--1458.
\newblock \doi{10.1140/epjst/e2015-02470-3}.

\bibitem[{Liao and Wang(2014)}]{LiaoW-2014}
Liao, S. and Wang, P. (2014).
\newblock On the mathematically reliable long-term simulation of chaotic
  solutions of {L}orenz equation in the interval [0,10000].
\newblock \emph{Science China Physics, Mechanics and Astronomy}, 57(2),
  330--335.

\bibitem[{Lorenz(1963)}]{Lorenz-1963}
Lorenz, E. (1963).
\newblock Deterministic nonperiodic flow.
\newblock \emph{J. Atmos. Sci.}, 20(2), 130--141.

\bibitem[{Mendes and Mendes(2005)}]{MendesMen:2-2005}
Mendes, D. and Mendes, V. (2005).
\newblock Control of chaotic dynamics in an {O}{L}{G} economic model.
\newblock \emph{Journal of Physics: Conference Series}, 23, 158--181.

\bibitem[{Naimzada and Pireddu(2015)}]{NaimzadaP:6-2015}
Naimzada, A. and Pireddu, M. (2015).
\newblock Introducing a price variation limiter mechanism into a behavioral
  financial market model.
\newblock \emph{Chaos}, 25, 83--112.

\bibitem[{Neck(2009)}]{Neck:15-2009}
Neck, R. (2009).
\newblock Control theory and economic policy: Balance and perspectives.
\newblock \emph{Annual Reviews in Control}, 33, 79--88.

\bibitem[{Orlando(2006)}]{Orlando:4-2006}
Orlando, G. (2006).
\newblock Routes to chaos in macroeconomic theory.
\newblock \emph{Journal of Economic Studies}, 33(6), 437--468.

\bibitem[{Ott et~al.(1990)Ott, Grebogi, and Yorke}]{OttGY-1990}
Ott, E., Grebogi, C., and Yorke, J. (1990).
\newblock Controlling chaos.
\newblock \emph{Physical review letters}, 64(11), 1196.

\bibitem[{Pchelintsev et~al.(2019)Pchelintsev, Polunovskiy, and
  Yukhanova}]{PchelintsevPY-2019-LorenzUpo}
Pchelintsev, A., Polunovskiy, A., and Yukhanova, I. (2019).
\newblock The harmonic balance method for finding approximate periodic
  solutions of the {L}orenz system.
\newblock \emph{Tambov University Reports. Series: Natural and Technical
  Sciences}, 24, 187--203.
\newblock (in Russian).

\bibitem[{Pyragas(1992)}]{Pyragas-1992}
Pyragas, K. (1992).
\newblock Continuous control of chaos by selfcontrolling feedback.
\newblock \emph{Phys. Lett. A.}, 170, 421--428.

\bibitem[{Pyragas(2001)}]{Pyragas-2001}
Pyragas, K. (2001).
\newblock Control of chaos via an unstable delayed feedback controller.
\newblock \emph{Phys. Rev. Lett.}, 86, 2265--2268.

\bibitem[{Rosser(2000)}]{Rosser:Book-2000}
Rosser, J. (2000).
\newblock \emph{From catastrophe to chaos: A general theory of economic
  discontinuities}, volume 1. Mathematics, Microeconomics, Macroeconomics, and
  Finance.
\newblock Springer.

\bibitem[{Salarieh and Alasty(2009)}]{SalariehAl:3-2009}
Salarieh, H. and Alasty, A. (2009).
\newblock Chaos control in an economic model via minimum entropy strategy.
\newblock \emph{Chaos, Solitons \& Fractals}, 40, 839--847.

\bibitem[{Shapovalov et~al.(2004)Shapovalov, Kablov, Bashmakov, and
  Avakumov}]{ShapovalovKBA-2004}
Shapovalov, V., Kablov, V., Bashmakov, V., and Avakumov, V. (2004).
\newblock \emph{Synergetics and problems in control theory}, chapter
  Sinergeticheskaya model ustoychivosty sredney firmy (in {R}ussian), 454--464.
\newblock FIZMATLIT.

\bibitem[{Shapovalov and Kazakov(2015)}]{ShapovalKaz-2015}
Shapovalov, V. and Kazakov, N. (2015).
\newblock The {L}orentz attractor and other attractors in the economic system
  of a firm.
\newblock \emph{Journal of Physics: Conference Series}, 574, \ {a}rt. num.
  012084.

\bibitem[{Thomas et~al.(2019)Thomas, Olufsen, Sepulchre, Iglesias, Ijspeert,
  and Srinivasan}]{ThomasOSIIS:med-2019}
Thomas, P., Olufsen, M., Sepulchre, R., Iglesias, P., Ijspeert, A., and
  Srinivasan, M. (2019).
\newblock Control theory in biology and medicine.
\newblock \emph{Biological Cybernetic}, 113(1-2), 1--6.

\bibitem[{Viswanath(2001)}]{Viswanath-2001}
Viswanath, D. (2001).
\newblock The {L}indstedt--{P}oincar\'{e} technique as an algorithm for
  computing periodic orbits.
\newblock \emph{SIAM review}, 43(3), 478--495.

\bibitem[{Wieland and Westerhoff(2005)}]{WielandW:10-2005}
Wieland, C. and Westerhoff, F. (2005).
\newblock Exchange rate dynamics, central bank interventions and chaos control
  methods.
\newblock \emph{Journal of Economic Behavior \& Organization}, 58, 117--132.

\bibitem[{Xu et~al.(2002)Xu, Li, Bishop, and Galvanetto}]{XuBG:18-2002}
Xu, D., Li, Z., Bishop, S.R., and Galvanetto, U. (2002).
\newblock Estimation of periodic-like motions of chaotic evolutions using
  detected unstable periodic patterns.
\newblock \emph{Pattern Recognition Letters}, 23(1-3), 245--252.

\bibitem[{Yu et~al.(2012)Yu, Cai, and Li}]{YuCLi:12-2012}
Yu, H., Cai, G., and Li, Y. (2012).
\newblock Dynamic analysis and control of a new hyperchaotic finance system.
\newblock \emph{Nonlinear Dynamics}, 67(3), 2171--2182.

\end{thebibliography}


\end{document}